\documentstyle[epsfig]{article}
\def\etal{{\it et~al.}}
\def\oyobi{\&}

\title{ 
General Relativistic Simulations of Jet \\
Formation by a Rapidly Rotating Black Hole
}

\author{Shinji Koide \\
Faculty of Engineering, Toyama University,\\
Gofuku 3190, Toyama 930-8555, \\
Japan. \\
\and
David L. Meier \\
Jet Propulsion Laboratory, \\
4800 Oak Grove Dr. Pasadena, CA 91109,\\
USA.\\
\and
Kazunari Shibata\thanks{Present address: Hanayama Observatory,
Kyoto University, Kita-Hanayama, Yamashina-ku, Kyoto, 607-8471, Japan.}, 
Takahiro Kudoh \\
National Astronomical Observatory,\\
Mitaka, Tokyo 181-8588,\\
Japan.}

\begin{document}

\maketitle

\abstract{
    Recent observations of Galactic Black Hole Candidates (BHCs)
    suggest that those that are superluminal jet sources have more
    rapid black hole spin rates than otherwise normal BHCs.  This
    provides observational support for models of
    relativistic jet formation that extract rotational energy of the
    central black hole.  To investigate this mechanism, we have
    developed a new general relativistic magnetohydrodynamic code in
    Kerr geometry.  Here we report on the first numerical simulation of
    the formation of a relativistic jet in a rapidly-rotating
    ($a=0.95$) Kerr black hole magnetosphere.  
    We assume that the initial velocity of the disk is zero.
    We find that the maximum
    velocity of the jet reaches $0.93c$ (Lorentz factor, $2.7$) and
    the terminal velocity of the jet is $0.85c$ (Lorentz factor, $1.9$).
    On the other hand, for a non-rotating ($a=0$) Schwarzschild black
    hole, the maximum jet velocity is less than $0.6c$ as far as
    we calculated with the similar condition of the magnetosphere
    to that of the Kerr black hole case.
    These numerical results show the importance of the rapidly
    rotating black hole for the relativistic jet formation clearly.
%    Based on these
%    results, we propose a new model of relativistic jet formation from
%    rotating black holes in our Galaxy and active galactic nuclei.
%    This model may also be applicable to gamma-ray bursts.
}

\section{\bf Introduction}

    Radio jets ejected from radio loud active galactic nuclei (AGNs)
    sometimes show proper motion with apparent velocity exceeding the
    speed of light $c$ \cite{pearson81,hughes91}.  
    The widely-accepted explanation for
    this phenomenon, called superluminal motion, is relativistic jet
    flow in a direction nearly along the observer's line-of-sight with
    a Lorentz factor greater than 2.0 \cite{rees66}.  
    Such relativistic motion is
    thought to originate from a region very close to the putative
    supermassive black hole which is thought to power each AGN 
    \cite{lindenbell69,rees84}.  
    On the other hand, the great majority of AGN are radio quiet and do not
    produce powerful relativistic radio jets \cite{rees84}.  
    These two classes of
    active objects (radio loud and quiet) are also found
    in the black hole candidates (BHCs) in our own Galaxy.  Objects
    with superluminal jets, such as GRS 1915+105 and GRO J1655-40,
    belong to the radio loud class \cite{mirabel94,tingay95}.  
    Other objects such as Cyg X-1 and GS 1124-68 are 
    relatively radio quiet and produce little or no jet.

    What causes the difference between the two classes?  Recent
    observations of the BHCs in our Galaxy suggest that the Galactic
    superluminal sources contain very rapidly rotating black holes 
    (specific angular momentum of rotating black hole $a
    = 0.9-0.95$), while the black holes in Cyg X-1 and GS 1124-68 are
    spinning much less rapidly ($a = 0.3-0.5$) \cite{cui98}.  According to
    recent (nonrelativistic) studies of magnetically-driven jets 
    from the accretion disk by Kudoh \oyobi\ Shibata 
    \cite{kudo95,kudo97a}, %(1995, 1997a), 
    the terminal velocity of the formed jet
    is comparable to the rotational velocity of the disk at the foot of
    the jet. Further nonrelativistic simulations of jet formation 
    confirm these results \cite{kudo97b,ouyed97}, 
    except for the extremely large magnetic field case \cite{meier97}
    in which very fast jets can be produced.
    The rotation velocity at the innermost stable orbit of the
    Schwarzschild black hole ($r= 3r_S$) is $0.5c$.  
    (Here $r_{\rm S} = 2 G M_{\rm BH}/c^2$ is the Schwarzschild 
    radius, where $G$ and $M_{\rm BH}$ are
    the gravitational constant and black hole mass, respectively.) In
    addition, recent studies indicate that the poloidal magnetic field
    strengths in disks around non-rotating black holes may be rather
    weak \cite{livio98}.
    Therefore, a jet produced by MHD acceleration from an accretion
    disk around a non-rotating black hole should be sub-relativistic
    and very weak.  In fact, numerical simulations of jet formation in
    a Schwarzschild metric show only sub-relativistic jet flow \cite{koide99},
    except for the case with the artificial condition of a hydrostatic 
    corona in equilibrium \cite{koide98}.

    Several mechanisms for relativistic jet formation from rotating
    black holes have been proposed 
    \cite{blandford77,punsly90,takahashi90,takahashi98}.  
    However, up until now no
    one has performed a self-consistent numerical simulation of the
    dynamic process of jet formation in a rotating black hole
    magnetosphere.  To this end, we have developed a Kerr
    general relativistic magnetohydrodynamic (KGRMHD) code.  In this
    paper we report briefly on what we believe is the first calculation
    of its kind --- simulation of relativistic jet formation in a
    rotating black hole magnetosphere.  When the rotation parameter
    of the rotating black hole $a = 0.95$, the maximum velocity
    of the formed jet reaches $0.93c$ (Lorentz factor, $\gamma = 2.7$),
    and the terminal velocity is $0.85c$ ($\gamma = 1.9$).  
    On the other hand, when $a$ vanishes, only a slow jet is
    formed even if the initial condition of magnetosphere is
    almost the same as that of the Kerr black hole case.
    This result clearly shows the importance of the rapidly spinning
    black hole for the formation of relativistic jet.
%    According to the numerical result, we propose a new model of
%    energy extraction of rotating black hole for
%    relativistic jet formation.

\section{\bf Numerical Method}

We use a 3 + 1 formalism of the general relativistic conservation
laws of particle number, momentum, and energy and Maxwell
equations with infinite electric conductivity \cite{thorne86}.
% (see Appendix C in \cite{koide99}).
The Kerr metric, which provides the spacetime around a rotating 
black hole, is used in the calculation.
When we use Boyer-Lindquist coordinates,
$x^0=ct$, $x^1=r$, $x^2=\theta$, and $x^3=\phi$, the Kerr
metric $g_{\mu \nu}$ is written as follows,
\begin{equation}
ds^2=g_{\mu \nu}dx^{\mu}dx^{\nu}
=-h_0^2(cdt)^2+\sum_{i=1}^{3} h_i^2(dx^i)^2-2h_3\Omega _3 cdt dx^3 .
\end{equation}

By modifying the lapse function in our Schwarzschild black hole code 
($\alpha = \root \of {1-r_{\rm S}/r}$) to be 
$\alpha = \root \of {h_0^2+\Omega _3^2}$, 
and adding some terms to $\Omega _3$, we were able to develop a 
KGRMHD code relatively easily. 
(See Appendix C in \cite{koide99} for more details on this procedure 
and the meaning of symbols used.) 

We use the Zero Angular Momentum Observer (ZAMO) system
for the 3-vector quantities, such as velocity ${\bf v}$, magnetic field
${\bf B}$, and so on.
For scalars, we use the frame comoving with the fluid flow.
The simulation of the Kerr (Schwarzschild) black hole is performed
in the region $0.75r_{\rm S}$ $(1.2r_{\rm S}) \leq r \leq
20r_{\rm S}$, $0 \leq \theta \leq \pi /2$ with $210 \times 70$
mesh points, assuming axisymmetry with respect to the $z$-axis
and mirror symmetry with respect to the plane $z=0$.
A simplified radiative boundary condition is employed
at $r=0.75 r_{\rm S}$ ($r=1.2 r_{\rm S}$) and $r=20r_{\rm S}$
for the Kerr (Schwarzschild) black hole case.
In the simulations, we use simplified tortoise coordinates,
$x={\rm log}(r/r_{\rm H} -1)$, where $r_{\rm H}$ is the 
radius of the black hole horizon. Here $r_{\rm H} =0.656 r_{\rm S}$
($r_{\rm H} = r_{\rm S}$) when $a=0.95$ ($a=0$). 
To avoid numerical oscillations, we use
a simplified TVD method \cite{koide99,davis84,koide96,koide97}.
We checked the KGRMHD code by computing Keplerian motion around
a rotating black hole and comparing with analytic results.

\section{\bf Results}

\begin{figure}
\hspace{-1cm}
\epsfig{figure=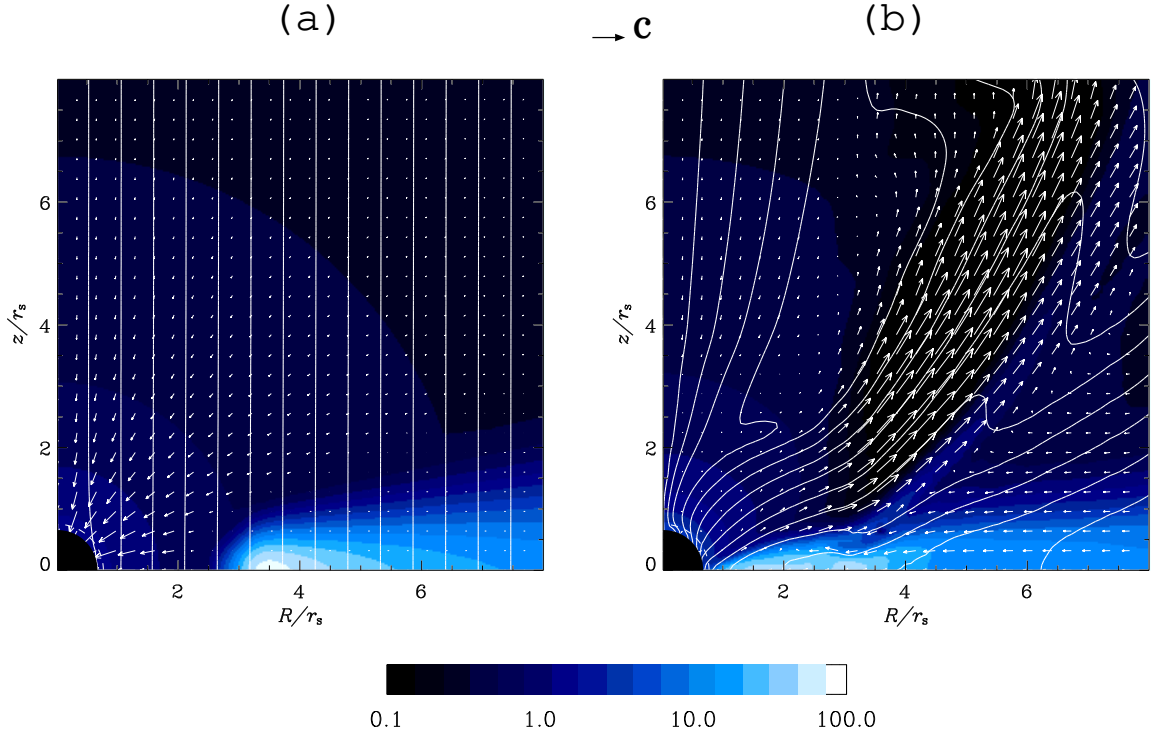}
\caption{
A relativistic jet formation around a Kerr black hole
(rotation parameter $a=0.95$). (a) Initial. (b) Final stage
at $t=65 \tau _{\rm S}$. Color shows the logarithm of mass density. 
Vector indicates the velocity. 
% Near the black hole horizon, falling velocity is $c$ at 
% the initial stage.
The lines show the magnetic field. 
The broken line near the horizon shows the inner boundary
of the calculation region.
The relativistic jet
are formed at $t=65 \tau _{\rm S}$. The maximum velocity of the
jet is $0.93c$ (Lorentz factor, $\gamma = 2.7$) and the
terminal velocity is $0.85c$ ($\gamma = 1.9$).
The solid straight line shows the plot line used in Fig. 3.
}
\end{figure}
Figure 1 shows the initial and final states of the calculation
of relativistic jet formation near a Kerr black hole.
These figures show the rest mass density (color), velocity
(vector), and magnetic field (solid lines) in
$0 \leq R \leq 8 r_{\rm S}$, $0 \leq z \leq 8 r_{\rm S}$.
The poloidal magnetic field lines are drawn as the contour lines
of the azimuthal  ($\phi$) component of the four vector potential
$(A_t, A_r, A_{\theta}, A_{\phi})$.
The black region at the origin shows the black hole horizon.
The specific angular momentum of the black hole is $a=0.95$
and the radius is $r_{\rm H} =0.656 r_{\rm S}$. 
The initial state in the simulation consists of
a hot corona and a cold accretion disk around the black hole (Fig. 1a).
In the corona, plasma are assumed to be nearly
stationary falling with the specific enthalpy
$h/\rho c^2 =1+\Gamma p/[(\Gamma -1)\rho c^2] =1.3$,
where $\rho$ is the rest mass density, $p$ is the pressure, and
$\Gamma$ is specific heat ratio and set $\Gamma =5/3$. 
Far from the hole,
it becomes the stationary transonic solution exactly.
The accretion disk is located at $|{\rm cot} \theta | \leq 0.125$,
$r \geq r_{\rm D} =3r_{\rm S}$ and the initial
velocity of the disk is assumed to be zero.
The mass density of the disk is 300 times that of the
corona. In addition, the magnetic field crosses the accretion disk
perpendicularly. We use the vector potential $A_{\rm \phi}$
of the Wald \cite{wald74} solution to set the magnetic field, 
which produces a uniform magnetic field far from the Kerr black hole. 
However, we do not use $A_t$ from the Wald solution; 
instead, we use the ideal MHD condition ${\bf E} +{\bf v} \times {\bf B} =0$
to determine the electric field ${\bf E}$. 
Here the Alfv\'{e}n velocity and plasma beta value at the disk 
($r=3r_{\rm S}$) are $v_{\rm A} = 0.01c$ and $\beta \sim 12$,
respectively.

Figure 1b shows the state at $t= 65\tau _{\rm S}$, 
where $\tau _{\rm S}$ is defined as $\tau _{\rm S} \equiv r_{\rm S}/c$.
The accretion disk falls into the black hole rapidly, and
the disk plasma crosses the horizon,
as suggested by the crowded magnetic field lines near the horizon.
The magnetic field lines become radial due to the dragging by the disk
near the black hole.
% The existence of the magnetized heavy disk is important
% to form the relativistic jet powered by the rapidly rotating
% black hole as we discuss latter. 
The jet is formed almost along the magnetic field lines, while
outside of the jet, the magnetic field lines
are bent strongly.  The maximum velocity of the jet
reaches to $0.93c$ (Lorentz factor, $\gamma = 2.7$) at 
$R=5.3r_{\rm S}$, $z=4.4r_{\rm S}$ and its terminal velocity
is $0.85c$ ($\gamma = 1.9$).
The terminal velocity is defined by the jet velocity far from 
the black hole ($r \sim 10 r_{\rm S}$).
We found the jet has a two-layered-shell structure with 
respect to the density. The low density part of the jet is fast and
relativistic, while the high density part at the outer 
shell is slower and sub-relativistic.
The fast jet dominates the jet and the slow, dense jet is thin.
% The fast, low density jet is accelerated due to the extraction
% of the rotational energy of the spinning black hole
% as will be discussed later.
The high density jet is connected to the high density
region of the accretion disk.
% The magnetic tension of the bent magnetic field and high pressure
% cause the acceleration of the slow, high
% density plasma from the accretion disk.
\begin{figure}
\hspace{-1cm}
\epsfig{figure=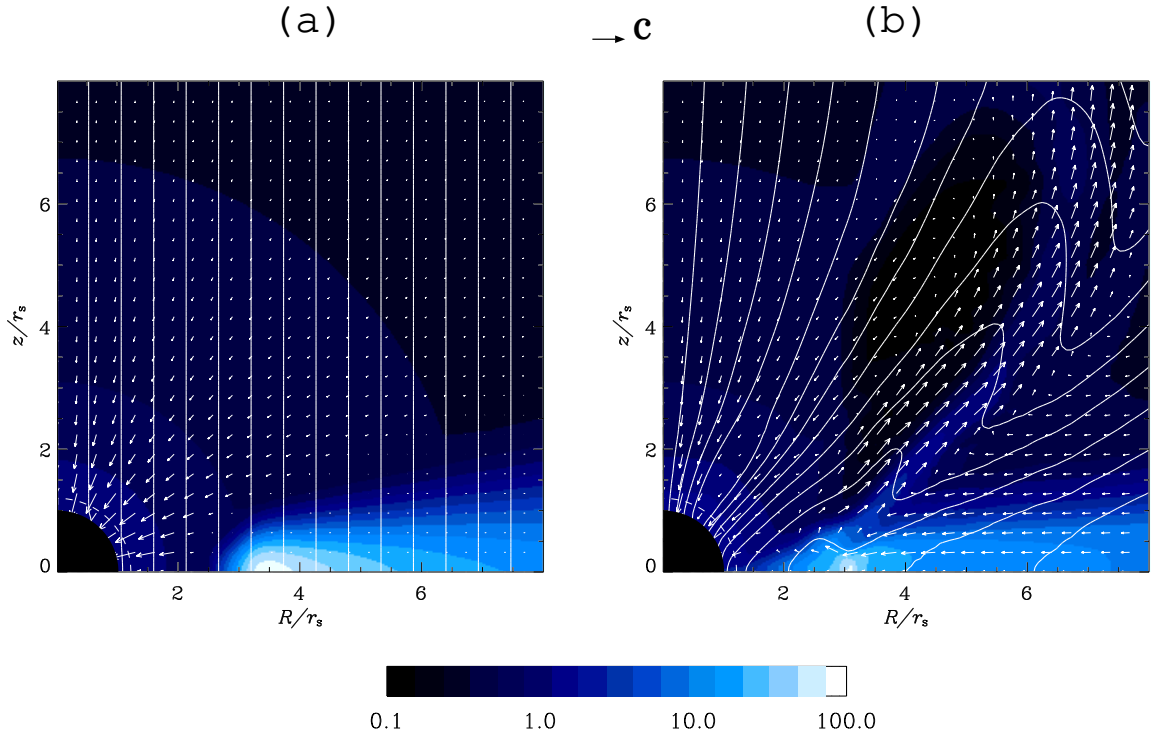}
\caption{
A sub-relativistic jet formation around Schwarzschild black hole
($a=0$). (a) Initial (b) Final stage at $t=65 \tau _{\rm S}$.
Color shows the logarithm of mass density. Vector indicates
the velocity. The lines show the magnetic field.
The broken line near the horizon shows the inner boundary
of the calculation region.
The maximum and terminal velocities of the jet are
$0.6c$ and $0.45c$, both sub-relativistic, respectively.
The solid straight line shows the plot line used in Fig. 4.
}
\end{figure}

To show the importance of the black hole rotation for 
relativistic jet formation, we also performed a simulation of 
the Schwarzschild black hole case ($a=0$) (see Fig. 2).
The initial condition is almost the same as that of Kerr black
hole case except for the black hole rotation (Fig. 2a).
In the initial stage, the coronal plasma is assumed to be
stationary falling with $h/\rho c^2 = 1.3$ (see Fig. 2a).
The disk is located at the same place of the Kerr black hole case:
$|{\rm cot} \theta | \leq 0.125$, $r \geq r_{\rm D} =3r_{\rm S}$,
and the density of the disk is also 300 times larger than that of corona.
The uniform magnetic field crosses the disk perpendicularly.
The strength is the same as that of the Kerr case.
Figure 2b shows the state at $t=65 \tau _{\rm S}$.
The maximum velocity of the jet is $0.6c$ and
the jet terminal velocity is $0.45c$, both sub-relativistic.
The jet also has two-layered-shell structure. The inner shell has
low density, while the outer shell has high density. 
The high density jet dominates the jet, which is different
from the Kerr black hole case. 
The disk plasma also falls and crosses the event horizon
rapidly as the Kerr black hole case. 
Note that the inner edge of the high density region is not 
identical with that of disk plasma.
The plasma of the disk passes through the edge and 
falls into the black hole. 
The small high pressure region is formed at $R \sim 3 r_{\rm S}$
in the disk. 
The high pressure is caused by the shock in front of the region.
% At $R \sim 3 r_{\rm S}$,
% a shock is located and gas pressure increases strongly
% below the shock. The ballooning disk ($2.5 r_{\rm S} \leq
% R \leq 3 r_{\rm S}$) is caused by the high pressure.
The dense jet is accelerated by the high pressure.
This mechanism is the same as that of the case of the
Kepler disk around the Schwarzschild black hole \cite{koide98}.
The magnetic field lines are also bent by the jet strongly.

\begin{figure}
\begin{center}
\epsfig{figure=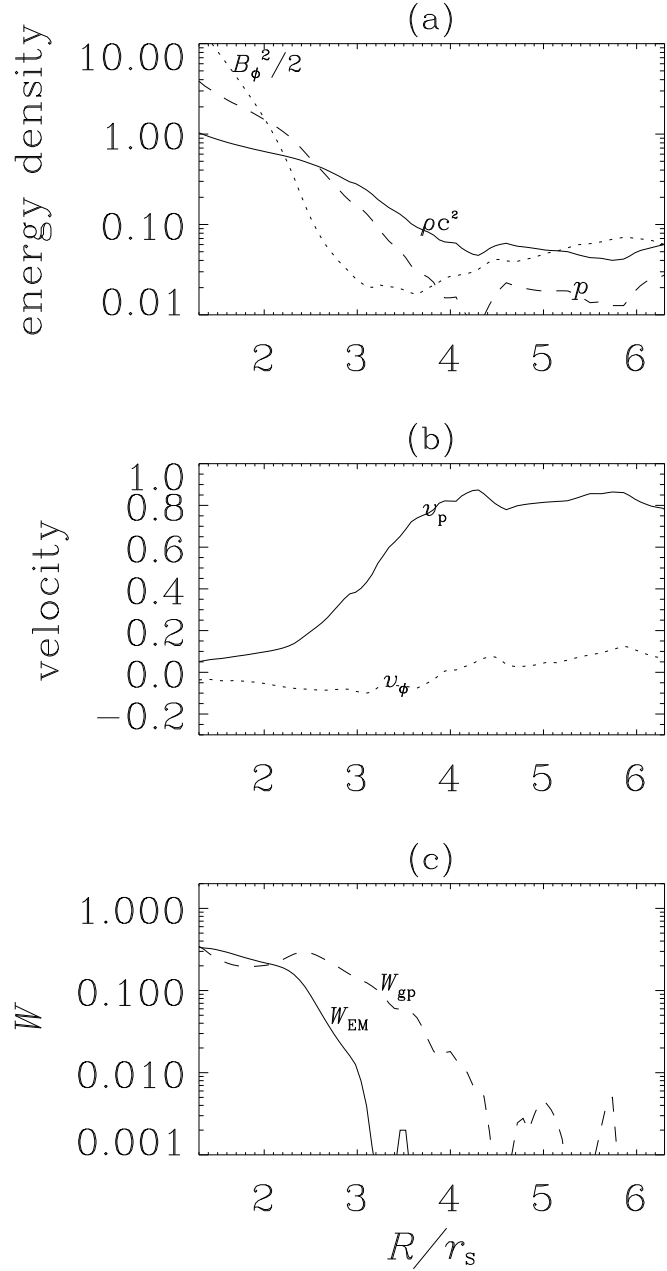}
\caption{
Physical quantities of the Kerr black hole case along $z= R -0.4r_{\rm S}$
$(1.3r_{\rm S} \leq R \leq 6.3 r_{\rm S}$) at $t=65 \tau _{\rm S}$
(see straight line in Fig. 1b).
(a) Mass density, $\rho$, gas pressure, $p$, and poloidal magnetic 
pressure, $B_{\phi}^2/2$.
(b) Poloidal velocity, $v_{\rm p}$ and azimuthal
velocity,  $v_{\phi}$.
(c) Power of electromagnetic field ($W_{\rm EM}$) and
gas pressure ($W_{\rm gp}$) to the acceleration of the jet.
}
\end{center}
\end{figure}

To show the structure of the relativistic jet
in the Kerr black hole case,
various physical quantities on the line $z=R - 0.4r_{\rm S}$ $(1.3r_{\rm S}
\leq R \leq 6.3r_{\rm S}$) are shown in Fig. 3. 
The line is almost along the fast, low density jet
(see straight line in Fig. 1b).
Figure 3a shows the magnetic pressure from the toroidal component
of the magnetic field, $B_{\phi}^2/2\mu$ is very large 
near the black hole ($1.3r_{\rm S} \leq R \leq 2.5 r_{\rm S}$)
where $\mu$ is the magnetic permeability of the vacuum.
Because initial magnetic field is 
$B_\phi =0$, $B_z =0.15\rho _0 c^2 \root \of {\mu}$,
where $\rho _0$ is the unit mass density, the field
increases to more than 60 times the initial magnetic field.
This amplification is caused by the shear of the plasma flow
observed by the Boyer-Lindquist frame or local nonrotating frame (LNRF)
due to the frame dragging effect of the rotating black hole
\cite{yokosawa91,yokosawa93,meier99}. 
This effect is called a {\it metric-shear-driven MHD dynamo}
by Meier \cite{meier99}.
% When we assume azimuthal velocity of the plasma is zero, 
With the assumption that the azimuthal velocity component
is zero, general relativistic Faraday law of induction and ideal MHD
condition yield,
\begin{equation}
\frac{\partial B_\phi}{\partial t} = f_1 B_r + f_2 B_\theta ,
\label{eqdyn}
\end{equation} 
where $f_1 = c(h_3/h_1) \partial (\Omega _3/h_3)/\partial r$,
and $f_2 = c(h_3/h_2) \partial (\Omega _3/h_3)/\partial \theta$.
Note that $f_2$ is one order smaller than $f_1$ when $a \sim 1$.
The ridge of the $f_1$ profile is located near the spherical
surface, $r = r_{\rm S}$
and the magnetic pressure has a maximum near the surface.
Where does the energy of the amplification of the magnetic field
comes from?
The energy does not come from the gravitational energy
or thermal energy of the disk, because this effect occurs even
when the plasmas of the disk and corona are rest and cool.
The only other possible energy source is the rotation of the
black hole itself. In fact, the increase in the azimuthal magnetic field
component depends on the shear of the rotational variable $\Omega _3$.
We conclude that the energy of the amplification of the magnetic
field is supplied by the extraction of the rotational energy
of the black hole. Note that this mechanism of the extraction of
the rotational energy is different from the previous
electromagnetic mechanisms \cite{blandford77,takahashi90}.
The detail comparison with these models will be presented
in our following paper.
If there is no disk, the magnetic tension causes
the azimuthal motion of the plasma to prevent the dynamo.
The heavy disk plays an important role to keep the
magnetic energy from the energy of the black hole rotation
to produce a relativistic jet.
% The very strong magnetic pressure $B_\phi ^2/2\mu$ pushes
% the plasma against the strong gravity of the black hole.
We estimate the power from the electro-magnetic field,
$W_{\rm EM}={\bf v} \cdot ({\bf E} + {\bf J} \times {\bf B}$)
and the gas pressure, $W_{\rm gp} = - {\bf v} \cdot \nabla p$
(Fig. 3c).
% Near the black hole ($1.3 r_{\rm S} \leq R \leq 2.0r_{\rm S}$), 
% the power of the electromagnetic field
% is dominant due to the very strong magnetic pressure
% $B_\phi ^2/2$. At $R \sim 2.5r_{\rm S}$, 
The power contributions of the electromagnetic field and gas 
pressure are comparable. The high gas pressure is caused by the 
shock in the disk at $r \sim 2 r_{\rm S}$.
The density and the pressure are small in the fast jet region
($3.5r_{\rm S} \leq R$), which means that
the jet comes from the corona, not the disk. 
Both the strong magnetic and gas pressures push the low density
plasma to become the relativistic jet.
Figure 3b shows the azimuthal velocity is small over the whole range. 
This indicates the jet acceleration is caused by 
the magnetic and gas pressure
rather than the magnetic tension (or centrifugal force).
The magnetically driven part of the mechanism is the same as that of 
Uchida \oyobi\ Shibta \cite{uchida85} and Shibata \oyobi\ Uchida 
\cite{shibata86}.
Punsly \oyobi\ Coroniti (1990) proposed an alternative
magnetic force driven mechanism of the relativistic jet
formation. They considered the magnetic tension
causes the azimuthal velocity to produce the jet due to
centrifugal force like the model of non-relativistic 
jet formation from the accretion disk \cite{blandford82}.
% Our model is complementary to their model and supported
% by the numerical calculation.
Figure 3b shows that in the fast jet region ($4r_{\rm S} \leq R$), 
the azimuthal velocity is small but positive ($v_{\phi} > 0$), 
which is remnant of the frame dragging by the rotating black hole.
In the acceleration region ($R < 4r_{\rm S}$), on the other hand, 
the azimuthal velocity is negative, as predicted by Meier 
\cite{meier99}.  Accretion of this negative angular momentum material 
ultimately reduces the spin of the black hole and is at least 
partially responsible for extracting the hole's rotational energy.
%This means that the acceleration suggested by Punsly \oyobi\ Coroniti
%(1990) also contributes to the acceleration of the jet but weakly. 
% The azimuthal velocity in the inner
% region ($R < 4r_{\rm S}$) is negative because 
% it is thought to be a part of Alfv\'{e}n wave.

\begin{figure}
\begin{center}
\epsfig{figure=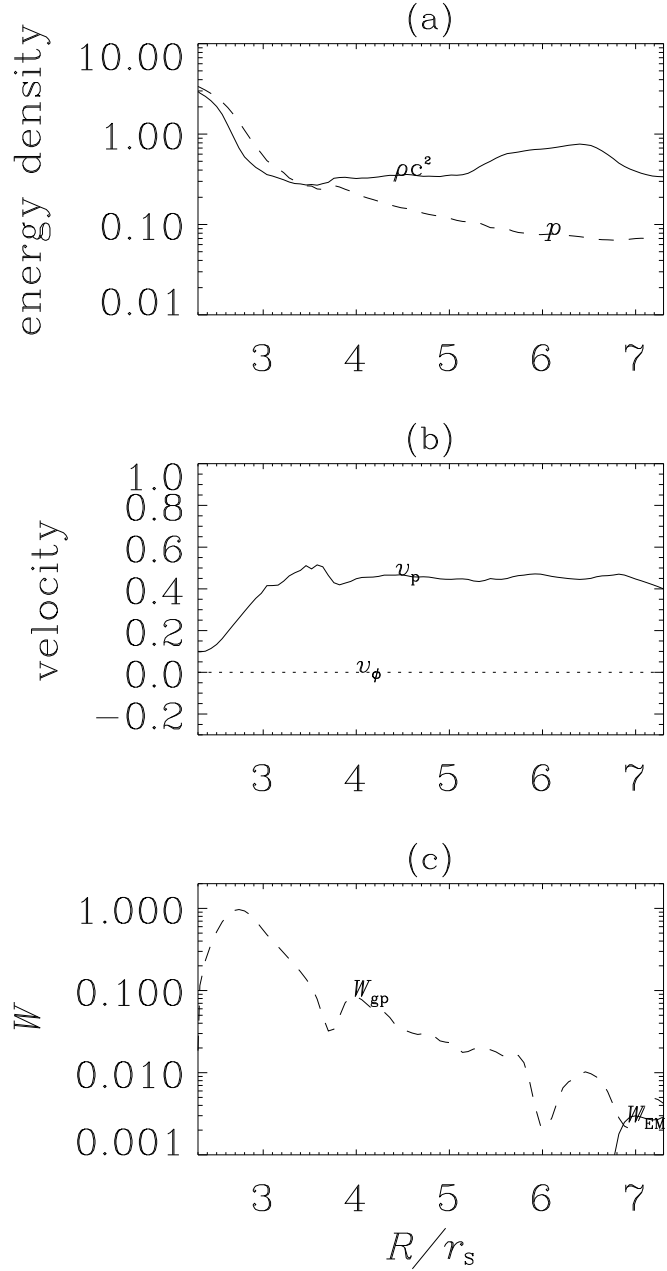}
\caption{
Physical quantities of the Schwarzschild black hole case
along $z= R -2r_{\rm S}$ $(2.3r_{\rm S} \leq
R \leq 7.3 r_{\rm S}$) at $t=65 \tau _{\rm S}$
(see straight line in Fig. 2b).
(a) Mass density, $\rho$, gas pressure, $p$, and poloidal magnetic 
pressure, $B_{\phi}^2/2$.
(b) Poloidal velocity, $v_{\rm p}$ and azimuthal
velocity, $v_{\phi}$.
(c) Power of electromagnetic field ($W_{\rm EM}$) and
gas pressure ($W_{\rm gp}$) to the acceleration of the jet.
}
\end{center}
\end{figure}
Figure 4 shows the physical variables of the Schwarzschild case
at $t=65 \tau _{\rm S}$ on the line $z=R - 2 r_{\rm S}$
$(2.3 r_{\rm S} \leq R \leq 7.3r_{\rm S}$).
The line is also almost along the main, dense jet (see straight
line in Fig. 2b).
The terminal velocity of the jet is $0.45c$, 
sub-relativistic (Fig. 4b).
However, the mass density and pressure are relatively high (Fig. 4a).
This means that the jet comes from the disk.
The jet is accelerated only by gas pressure (Fig. 4c).
The high pressure is caused by the shock heating in the disk
as the same as in the case of a Keplerian disk around
a Schwarzschild black hole \cite{koide98}.
It is reasonable that the azimuthal components of velocity
and magnetic field vanish because both the black hole and
disk have no rotation. In such a case, magnetic acceleration
is not permitted and only a gas pressure driven jet
may be formed.

\section{\bf Discussion}

We have also performed a maximally rotating black hole case
($a=1$). Preliminary results show that a strong
relativistic jet is formed, with a maximum velocity of $0.98c$
(Lorentz factor 5!) and a terminal velocity of $0.9c$
(Lorentz factor, $2.3$).
Unfortunately, the inner boundary of this calculation is located
at $r = 0.8 r_{\rm S}$, which is a little bit far from the
horizon, $r_{\rm H} = 0.5 r_{\rm S}$.
We believe that a faster relativistic jet will be formed when
we use an inner boundary nearer to the horizon. 
On the other hand, in the Schwarzschild case
with similar initial conditions in the magnetosphere, 
only a sub-relativistic jet is formed. We have also performed
a number of simulations with rotating disks around Schwarzschild
black holes \cite{koide99}. In these cases, all formed jets are 
sub-relativistic, except for a special case with hydrostatic
equilibrium corona \cite{koide98}. 

Of course, more simulations of jets formed by rotating black 
holes accreting magnetized plasma need to be done in order to 
fully understand this formation mechanism. 
However, our numerical results show clearly that the spin of the 
black hole is important in the formation a relativistic jet.
A comprehensive model of relativistic jet formation
around Kerr black hole, along with more detailed 
data analysis, will be presented in our following paper.

Recent X-ray observations of black hole candidates
(BHCs) in our Galaxy show the relationship between the jet
speed and the rotation of the central objects \cite{cui98}.
A quasi-periodic-oscillation (QPO) is found in the power 
spectrum of the Galactic superluminal source GRO J1655-40
\cite{remillard97} and GRS 1915+105 \cite{morgan97}.
Using QPO observations, it is concluded that the both Galactic
superluminal sources contain very rapidly rotating black holes
($a \sim 0.95$). Similarly, using the QPOs in Cyg X-1 \cite{cui97}
and in GS 1124-68 \cite{belloni97b}, it is shown that
the black holes in these two sources are spinning less rapidly
($a \sim 0.35$ and $a \sim 0.48$, respectively). Based on these
results, Cui, Zhang, \oyobi\ Chen (1998) proposed that the difference between
BHCs that are also superluminal jet sources and otherwise {\it normal} 
BHCs is in the spin of their black holes. This observational
result is consistent with our numerical results.

\vspace{0.5cm} 

% We thank K.-I. Nishikawa for his discussion and kind help.
It is a pleasure to thank K.-I. Nishikawa for suggestions that
significantly improved the manuscript. 
One of author (S. K.) thanks M. Inda-Koide for her discussion 
and important comments for this study. 
We also thank M. Takahashi, A. Tomimatsu, J. F. Hawley,
R. D. Blandford, M. C. Begelman, P. E. Hardee,
J.-I. Sakai, and R. L. Mutel 
for their discussions and encouragement. 
We appreciate the support of the National Institute for Fusion Science
and the National Astronomical Observatory
for the use of super-computers.


\newpage

\begin{thebibliography}{99}
\bibitem[1]{pearson81} 
J. J. Pearson, \etal\ Nature {\bf 290} (1981) 365.

\bibitem[2]{hughes91} 
P. A. Hughes, eds., {\it Beams and Jets in 
Astrophysics} (Cambridge University Press, New York, 1991).

\bibitem[3]{rees66} 
M. J. Rees, Nature {\bf 211} (1966) 468. %-470.

\bibitem[4]{lindenbell69}
D. Linden-Bell, Nature {\bf 223} (1969) 690.

\bibitem[5]{rees84} 
M. J. Rees, Ann. Rev. Astron. Ap. {\bf 22} (1984) 471. %-506.

\bibitem[6]{mirabel94} 
I. F. Mirabel \oyobi\ L. F. Rodriguez, 
Nature {\bf 371} (1994) 46. %-48.

\bibitem[7]{tingay95} 
S. J. Tingay, \etal\ Nature {\bf 374} (1995) 141.

\bibitem[8]{cui98}
W. Cui, S. N. Zhang, \oyobi\ W. Chen, 
ApJ {\bf 484} (1998) 383.

\bibitem[9]{kudo95} 
T. Kudoh \oyobi\ K. Shibata, 
ApJ {\bf 452} (1995) L41. 

\bibitem[10]{kudo97a} 
T. Kudoh \oyobi\ K. Shibata, 
ApJ {\bf 476} (1997) 632.

\bibitem[11]{kudo97b}
T. Kudoh \oyobi\ K. Shibata, ApJ {\bf 474} (1997) 362.

\bibitem[12]{ouyed97} 
R. Ouyed, R. E. Pudritz, \oyobi\ J. M.  Stone, 
Nature {\bf 385} (1997) 409. %-414.
      
\bibitem[13]{meier97} 
D. L. Meier, S. Edgington, P. Godon, D. G. Payne, \oyobi\ K. R. Lind,
Nature {\bf 388} (1997) 350.

\bibitem[14]{livio98} 
M Livio, K. W. Ogilvie, and J. E. Pringle, ApJ, in press.

\bibitem[15]{koide99} 
S. Koide, K. Shibata, \oyobi\ T. Kudoh, ApJ, in press.

\bibitem[16]{koide98} 
S. Koide, K. Shibata, \oyobi\ T. Kudoh, ApJ {\bf 495} (1998) L63. %-L66.

\bibitem[17]{blandford77} 
R. D. Blandford \oyobi\ R. Znajek, MNRAS
{\bf 179} (1977) 433.

\bibitem[18]{punsly90} 
B. Punsly \oyobi\ F. V. Coroniti, ApJ {\bf 354} (1990) 583.

\bibitem[19]{takahashi90} 
M. Takahashi, S. Nitta, Y. Tatematsu, \oyobi\ A. Tomimatsu, 
ApJ {\bf 363} (1990) 206.

\bibitem[20]{takahashi98} 
M. Takahashi \oyobi\ S. Shibata, PASJ {\bf 50} (1998) 271.

\bibitem[21]{thorne86}
K. S. Thorne, R. H. Price, \oyobi\ D. A. Macdonald, 
{\it Membrane Paradigm} (Yale University Press, New Haven and 
London, 1986).

\bibitem[22]{davis84}
S. F. Davis, NASA Contractor Rep. 172373,
ICASE Rep. No. 84-20 (1984).

\bibitem[23]{koide96} 
S. Koide, K.-I. Nishikawa, \oyobi\ R. L. Mutel, 
ApJ {\bf 463} (1996) L71. %-L74.

\bibitem[24]{koide97} 
S. Koide, ApJ {\bf 478} (1997) 66. %-69. 

\bibitem[25]{wald74}
R. M. Wald, Phys. Rev. D {\bf 10} (1974) 1680. %-1685.

\bibitem[26]{yokosawa91}
M. Yokosawa, T. Ishizuka, \oyobi\ Y. Yabuki, 
PASJ {\bf 43} (1991) 427.

\bibitem[27]{yokosawa93}
M. Yokosawa, PASJ {\bf 45} (1993) 207.

\bibitem[28]{meier99}
D. L. Meier, ApJ, submitted.

\bibitem[29]{uchida85} 
Y. Uchida \oyobi\ K. Shibata, 
PASJ {\bf 37} (1985) 515. %-535.

\bibitem[30]{shibata86} 
K. Shibata \oyobi\ Y. Uchida, 
PASJ {\bf 38} (1986) 631. %-660.

\bibitem[31]{blandford82} 
R. D. Blandford \oyobi\ D. G. Payne, MNRAS
{\bf 199} (1982) 883.

\bibitem[32]{remillard97}
R. A. Remillard, E. H. Morgan, J. E. McClintock,
C. D. Bailyn, A. Oroszek, \oyobi\ J. Greiner, 
in Proc. 18th Texas Symp. on Relativistic Astrophysics,
ed. A. Olinto, J. Frieman, \oyobi\ D. Schramm (Singapore:
World Scientific Press, 1999) in press.

\bibitem[33]{morgan97}
E. H. Morgan, R. A. Remillard, \oyobi\ J. Greiner, ApJ
{\bf 482} (1977) 993.

\bibitem[34]{cui97}
W. Cui, S. N. Zhang, W. Focke, \oyobi\ J. H. Swank, 
MNRAS {\bf 290} (1997) L65.

\bibitem[35]{belloni97b} 
T. Belloni, M. van der Klis, W. H. G. Lewin, J. van Paradijs,
T. Dotani, K. Mitsuda, \oyobi\ S. Miyamoto, A \oyobi\ A {\bf 322}
(1997) 857.

% \bibitem[Iwamoto \etal\ 1998]{iwamoto98}
% Iwamoto, K. \etal\ 1998, Nature {\bf 395}, 672.

% \bibitem[Kippen \etal\ 1998]{kippen98} 
% Kippen, R. M. \etal\ 1998, ApJ {\bf 506}, L27.

% \bibitem[Kulkarni \etal\ 1998]{kulkarni98} 
% Kulkarni, S. R., \etal\ 1998, Nature {\bf 395}, 663.
\end{thebibliography}
\end{document}